\def\bq{\begin{equation}}
\def\eq{\end{equation}}
\def\bqa{\begin{eqnarray}}
\def\eqa{\end{eqnarray}}
\def\bqb{\begin{eqnarray*}}
\def\eqb{\end{eqnarray*}}

\def\mz{M_Z^2}

\def\q{q^2}
\def\p{ {\cal P}}
\def\g{\gamma}

\def\pr#1#2#3{ Phys. Rev. ${\bf{#1}}$ (#2) #3 }
\def\prl#1#2#3{ Phys. Rev. Lett. ${\bf{#1}}$ (#2) #3 }

\documentstyle[12pt]{article}
\begin{document}
\thispagestyle{empty}

\vspace*{1cm}
\hspace {-0.8cm} PM/94-34 \\
\hspace {-0.8cm} September 94 \vspace {2cm}\\
%---------------------titre ---------------------------------------
\begin{center}
{\Large\bf  A Probe of New Resonant Structures with}
\vspace{0.5cm}\\
{\Large\bf  Four Fermion
Processes at a 1 TeV $e^+e^-$ Collider  }
    \vspace{1cm}  \\
%-----------------------------------------------------------------
 {\Large  J. Layssac, F.M. Renard }
  \\
 Physique
Math\'{e}matique et Th\'{e}orique,\\
CNRS-URA 768, Universit\'{e} Montpellier II \\
 F-34095 Montpellier Cedex 5
\vspace {1cm}\\
 {\Large C. Verzegnassi} \\
 Dipartimento di Fisica Teorica,
 Universit\`{a} di Trieste \\
 Strada Costiera, 11 Miramare, I-34014 Trieste, \\
 and INFN, Sezione di Trieste, Italy
\vspace {1.5cm}  \\
 {\bf Abstract}
\end{center}
\noindent
 Oblique contributions from vector particles that are strongly
coupled to the known gauge bosons are calculated for the case of
various observables at a 1 TeV $e^+e^-$ collider. Constraints
provided by LEP 1 results are taken into account in a model-independent
way through a dispersion relation technique. Depending on the
assumed theoretical properties, mass limits of several
TeV should be observable.

\setcounter{page}{0}
\newpage

In this paper we investigate the potential of a 1 TeV $e^+e^-$ collider
for the indirect search of Technicolour-like vector bosons. High
precision tests at LEP 1 have already set rather stringent bounds on
Technicolour models, following the original
proposal of Peskin and Takeuchi \cite{peskin}. They are based on the
effect to the quantity S defined in \cite{peskin}, that appear
through the so-called \cite{lynn} oblique parts of the one-loop
contributions. We here apply a general formalism established
in recent publications
\cite{strong}, \cite{fourf}, \cite{vertex}, which allows
to calculate the relevant oblique contributions
to a number of processes
in future higher energies $e^+e^-$ experiments.\par
The main idea is that
of expressing the various effects in the form of a once-subtracted
dispersion integral, and of fixing the necessary subtraction constants
by suitable model-independent LEP 1 results. In this way, we are led
to a compact "representation" of several observables which present two
main advantages. The first one is that it allows to express the New
Physics contributions through convergent integrals. The second one is
that LEP 1 constraints are automatically incorporated in the
expressions of the observables. For example, the cross section
for muon production at cm energy $\sqrt{q^2}$,
$ \sigma_\mu(\q)$, at one loop level takes the form
\bqa \sigma_\mu^{SE}(\q)&=&
{4\pi\q\over3}\biggm\{  [{\alpha(\mz)\over q^2}]^2[1+2D_\gamma(\q)]
+ {1\over(q^2-\mz)^2+\mz\Gamma^2_Z}\nonumber\\
 & &\null \bigm[{3\Gamma_l\over M_Z}\bigm]^2[ 1-2 D_Z(\q)
-{16 s^2_1v_1\over1-v^2_1}D_{\gamma Z(q^2)}] \biggm\} \eqa
Here $\Gamma_l$ is the leptonic $Z$ width,
 $\alpha(\mz)=[1\pm0.001]/128.87$ and
 \bqa D_\gamma
(\q)\equiv\Delta\alpha(\q)-\Delta\alpha(\mz)
=-{\q-\mz\over\pi}\ \p\int_0^\infty {ds\ Im\ F_\gamma(s)\over
    (s-\q)(s-\mz)} \eqa
  \bqa D_Z(\q)\equiv Re\ [I_Z(\q)-I_Z(\mz)]
 = {\q-\mz\over\pi}\ \p\int_0^\infty {ds\ s\ Im\ F_{ZZ}(s)\over
    (s-\q)(s-\mz)^2} \eqa
  \bqa D_{\gamma Z}
(\q)\equiv Re\ [\Delta\bar \kappa'(\q)-\Delta\bar \kappa'(\mz)]
  = {\q-\mz\over\pi}\ \p\int_0^\infty {ds\ Im\ F_{\kappa'}(s)\over
    (s-\q)(s-\mz)} \eqa
 $$ (F_\kappa'=c_1/s_1\  F_{Z\g}\ \  , \ \
 s^2_1 c^2_1={\pi\alpha\over\sqrt2 G_\mu\mz}\ \ ,
\ \ s^2_1=1-c^2_1\simeq
  0.217 \ \
,\ \ v_1=1-4s^2_1). $$
The Imaginary parts which appear in these expressions are
constructed from the
self-energies. For Technicolour models, they are separately
gauge-invariant.\par
Similar representations were established for several other observables
like forward-backward asymmetries, polarization asymmetries and ratios
of hadron to muon production. For each observable we finally obtain an
expression that include the full effect of the oblique correction at
one-loop in the form:
\bq O(\q)= c_0[1+c_{\gamma}D_{\gamma}(\q)+c_ZD_Z(\q)+c_{\gamma
Z}D_{\gamma Z}(\q)]   \eq
where the analytic expressions of the various coefficients have been
derived for each observable and computed numerically in
\cite{fourf}.\par
We have presently used this formalism to calculate the
possible effects of a pair of vector (V)
and axial vector (A) resonances strongly coupled to the photon
and to the $Z$. The parameters which enter the expressions of the
imaginary parts of the various spectral functions are the couplings
$F_{V,A}$ and the masses $M_{V,A}$  (assumed to be larger than
$\sqrt{q^2}$). We have treated two different theoretical models.\par
In model(I) we consider a Technicolour-like framework in which
we exploited
the validity of the
two Weinberg sum rules \cite{weinberg}.
We only
retain their very general consequence, i.e.
the positivity of S. In a zero-width
approximation (in the actual computation of the effects on the
observables we have used a finite width description of the V,A
resonances) one has:
\bq  S = 4\pi[{F^2_V \over M^2_V} - {F^2_A \over M^2_A}] = {4\pi F^2_{\pi}
\over M^2_V}[1+{M^2_V\over M^2_A}] \eq
The present experimental constraint on S is \cite{Sexp}:
\bq -0.9 \leq \ S\ \leq 0.4 \eq
In this first model only the  \underline{positive} upper bound
is effective.\par
In model (II) we release the constraints due to the Weinberg sum rules.
This choice has the
consequence of introducing one more degree of freedom. It eliminates
the theoretical relation between $F_V$ and $F_A$. As a consequence S
can now take negative values, and in addition the strength of
the ratios $F_V/M_V$ and $F_A/M_A$ is no more bounded. We decided to
consider the limiting case consisting in a "strongly interacting
regime" for which the value of the ratio  $F_V/M_V$ is equal to twice
the QCD value\par

\bq {F_V\over M_V}=2{f_\rho\over m_\rho}={1\over\sqrt{2\pi}} \eq

 Then ,
 for every choice of $F^2_V/M^2_V$,
$F^2_A/M^2_A$ is allowed to saturate both limits imposed by the
experimental bounds on $S$.\par

Assuming a certain accuracy
for the measurement of each observable, one accordingly obtains
the observability limit of the self-energy effect that is
translated in an upper bound on the
masses $M_{V,A}$ . For $1\ TeV$ $ e^+e^-$ collider the assumed
accuracies are of
a relative one percent for  $\sigma_\mu$, $A_{FB,\mu}$, $A_{LR,h}$
,$R^{(5)}$, two percent for  $R_{b,\mu}$ and five percent for
$A_{\tau}$.\par
Results are shown in Fig. 1,2 for both models respectively,
the different curves corresponding to the various observables, and the
shaded area to the combined overall mass bounds. In model (I) the
resulting bounds on $M_{V,A} $ are located in the $2\ TeV$ range, and
rather strongly correlated. The only hadronic observable which
contributes appreciably to the bound is $A_{LR,h}$ , that allows to
improve the pure leptonic result by about $200\ GeV$.\\
In model (II) the effect of releasing the validity of the Weinberg sun
rule is roughly that of increasing the bounds on $(M_V,M_A)$ from the
$2\ TeV$ region to the $4\ TeV$ region.\par
Compared to the results obtained in \cite{strong}, \cite{fourf},
an improvement by a
factor two is found as compared to the case of a $500\ GeV$ collider
and by a factor 6 to 8 as compared to the $LEP 2$
case.\par
The mass range of $M_V$ and $M_A$ which is explored in this
indirect way should be able to give a definite hint of the existence of
Technicolour-like resonances or of any other strongly
coupled vector boson.\par


\begin{thebibliography}{99}


\bibitem{peskin} M.E. Peskin and T. Takeuchi, \pr{D46}{1991}{381}.

\bibitem{lynn} B.W. Lynn, M.E. Peskin and R. Stuart in "Physics at
LEP", J. Ellis and R. Peccei eds., CERN 86-02 (1986), Vol 1.

\bibitem{strong} J. Layssac, F.M. Renard and C. Verzegnassi,
\pr{D48}{1993}{4037}.

\bibitem{fourf} J. Layssac, F.M. Renard and C. Verzegnassi,
\pr{D49}{1994}{R2143}.

\bibitem{vertex} J. Layssac, F.M. Renard and C. Verzegnassi,
\pr{D49}{1994}{3650}.

\bibitem{weinberg} S. Weinberg,\prl{18}{1967}{507}.

\bibitem{Sexp} We use the last values for $\epsilon_3$ given by
G.Altarelli, CERN-TH-7319/94, June 94.

\bibitem{TC} S.Weinberg, \pr{D13}{1976}{974}, {\bf D19} (1979)
1277;
 L. Susskind, \pr{D20}{1979}{2619};
 E. Fahri and L. Susskind, \pr{D20}{1979}{3404}.
\end{thebibliography}
 \end{document}